\documentclass[journal,comsoc]{IEEEtran}

\usepackage[active]{srcltx} 
\usepackage{graphicx}
\usepackage{epsfig}
\usepackage{balance}
\usepackage{cite}
\usepackage{amssymb}
\usepackage[cmex10]{amsmath}
\usepackage{amsfonts}
\usepackage{amsthm}
\usepackage[applemac]{inputenc} 
\usepackage{subfigure}
\usepackage{color}
\usepackage{mathptmx} 
\usepackage{times} 
\usepackage{amsmath} 
\usepackage{makecell}
\usepackage{algorithm}
\usepackage{algorithmic}

\usepackage{hyperref}
\hypersetup{
    unicode=false,          
    pdftoolbar=true,        
    pdfmenubar=true,        
    pdffitwindow=false,     
    pdftitle={},    
    pdfauthor={},     
    pdfsubject={},   
    pdfcreator={},   
    pdfproducer={},  
    pdfkeywords={}, 
    pdfnewwindow=true,      
    colorlinks=false,       
    linkcolor=red,          
    citecolor=green,        
    filecolor=magenta,      
    urlcolor=cyan           
}
\usepackage{blindtext}

\newcommand{\argmax}[1]{\underset{#1}{\operatorname{arg}\operatorname{max}}\;}
\newcommand{\argmin}[1]{\underset{#1}{\operatorname{arg}\operatorname{min}}\;}
\newcommand{\eqdef}{\stackrel{\triangle}{=}}

\begin{document}
\newtheorem{theo}{Theorem}
\newtheorem{prop}{Proposition}
\newtheorem{cor}{Corollary}
\newtheorem{lem}{Lemma}

\theoremstyle{remark}

\theoremstyle{definition}
\newtheorem{defin}{Definition}
\newtheorem{assum}{Assumption}
\newtheorem{rem}{Remark}
\newtheorem{ins}{Insight}
\newtheorem*{sched}{Scheduling Problem}

\title{Software-Defined Network Controlled Switching between Millimeter Wave and Terahertz Small Cells}
\author{Angela Sara Cacciapuoti,~\IEEEmembership{Senior Member,~IEEE}, Ramanathan Subramanian, Marcello Caleffi,~\IEEEmembership{Senior Member,~IEEE}, and Kaushik Roy Chowdhury,~\IEEEmembership{Senior Member,~IEEE}
\thanks{A. S. Cacciapuoti and M. Caleffi are with the Department of Electrical Engineering and Information Technologies, University of Naples Federico II, Naples, 80125 Italy (e-mail: angelasara.cacciapuoti@unina.it; marcello.caleffi@unina.it).}
\thanks{R. Subramanian and K.R. Chowdhury are with the Department of Electrical and Computer Engineering, Northeastern University, Boston, MA 02115 USA (e-mail: rsubramanian@coe.neu.edu; krc@ece.neu.edu).}
}

\maketitle

\begin{abstract}
Small cells are a cost-effective way to reliably expand network coverage and provide significantly increased capacity for end users. The ultra-high bandwidth available at millimeter (mmWave) and Terahertz (THz) frequencies can effectively realize short-range wireless access links in small cells enabling potential uses cases such as driver-less cars, data backhauling and ultra-high-definition infotainment services. This paper describes a new software defined network (SDN) framework for vehicles equipped with transceivers capable of dynamically switching between THz and mmWave bands. We present a novel SDN controlled admission policy that preferentially handoffs between the mmWave and THz small cells, accommodates asymmetric uplink/downlink traffic, performs error recovery and handles distinct link states that arise due to motion along practical vehicular paths. We then analytically derive the resulting capacity of such a small cell network by accounting for the channel characteristics unique to both these spectrum bands, relative distance and the contact times between a given transceiver pair. We then formulate the optimal procedure for scheduling multiple vehicles at a given infrastructure tower, with regards to practical road congestion scenarios. The search for the optimal schedule is shown to be a NP-hard problem. Hence, we design a computationally-feasible polynomial-time scheduling algorithm that runs at the SDN controller and compare its performance against the optimal procedure and random access. Additionally, we present a simulation-based case study for the use case of data center backhauling in Boston city to showcase the benefits of our approach.     
\end{abstract}


\section{Introduction}
\label{sec:1}
Small cell densification in urban areas are expected to allow for the much needed scaling in capacity and ubiquitous coverage~\cite{Andrews5G} and, has enabled a surge in small cell deployments. Outdoor small cell deployments are expected to proliferate starting 2017~\cite{SmallCellForum}. While this is advantageous, the spectrum scarcity and congestion problem in the sub-6 GHz bands remain. Small cells that can utilize the available massive spectrum bandwidth in the millimeter-wave (mmWave) (around $30-100\,\mathrm{GHz}$) and Terahertz (THz) (around $0.1-10\,\mathrm{THz}$) frequencies promise a paradigm shift, in realizing fiber-equivalent wireless links~\cite{PetrovTHz}, leading up to several Tbps of effective data transfer rates, and further, freeing up the lower bands for macrocell to small cell communications. However, propagation in mmWave and THz bands is limited by the severe pathloss and atmospheric absorption. To counteract the significant attenuation, and extend coverage, high directivity gain antennas are used. The high data rate, limited coverage and reduced interference are attractive features and make these bands an excellent candidate for small cells.

Reliable and continuous high bandwidth connectivity within the next generation of vehicles will enable driver-less cars with on-the-road infotainment services using bulk media downloads, ultra-fast massive data transfers towards data backhauling and city-scale traffic optimization realized by uploading massive high-rate sensor data to the cloud for processing. Google's self-driving car, for example, generates sensor data at the rate of 750 MBps~\cite{GoogleCar} and automated driving cars are expected to generate in the order of 1 TB of sensor data in a single trip~\cite{SasPredict}. The sensor data can be used to remotely monitor the current state and predict a potential breakdown of the vehicle. Another potential use case can be to have the vehicles' camera images along with the location information be sent to the cloud for automakers to build detailed and accurate maps~\cite{ToyotaMap}. Self-driving cars, which are limited in their sensing range, will greatly benefit from precise maps, downloaded say when connecting to infrastructure, that reflect recent updates to navigate urban areas or the highways. Note that upload/download of such data will demand high throughput but there is no real-time requirement.

The ability to achieve data transfer rates in the order of several gigabits-per-second is key to enable such applications, so far unattainable through state of the art dedicated short-range communication (DSRC) and 4G cellular communication~\cite{ChoiPDBH16}. Additionally, fiber-based backhauling required to connect the small cells at scale to the core network will pose serious deployment challenges in terms of deployment time and wiring expenditure. Wireless backhauling using mmWave links, considered as an alternative solution, will be difficult to come by in urban settings (with trees and buildings of varying heights) given the reduced likelihood of LoS propagation conditions. In that regard, vehicles serving as digital mules will reduce deployment costs of fiber-based backhauling solutions~\cite{Kanno,Greenberg08}. It is important to note that fiber is expensive and can become congested and using vehicles may aid in bulk transfer of delay-tolerant information between data centers~\cite{Chen11-Infocom}. Further, there are inherent advantages of using vehicles as mobile-data caches. The vehicles are likely to contain region-specific content that can increase localized hits~\cite{Vigneri}. 

\begin{figure}[!t]
\centering
\includegraphics[width=1\columnwidth]{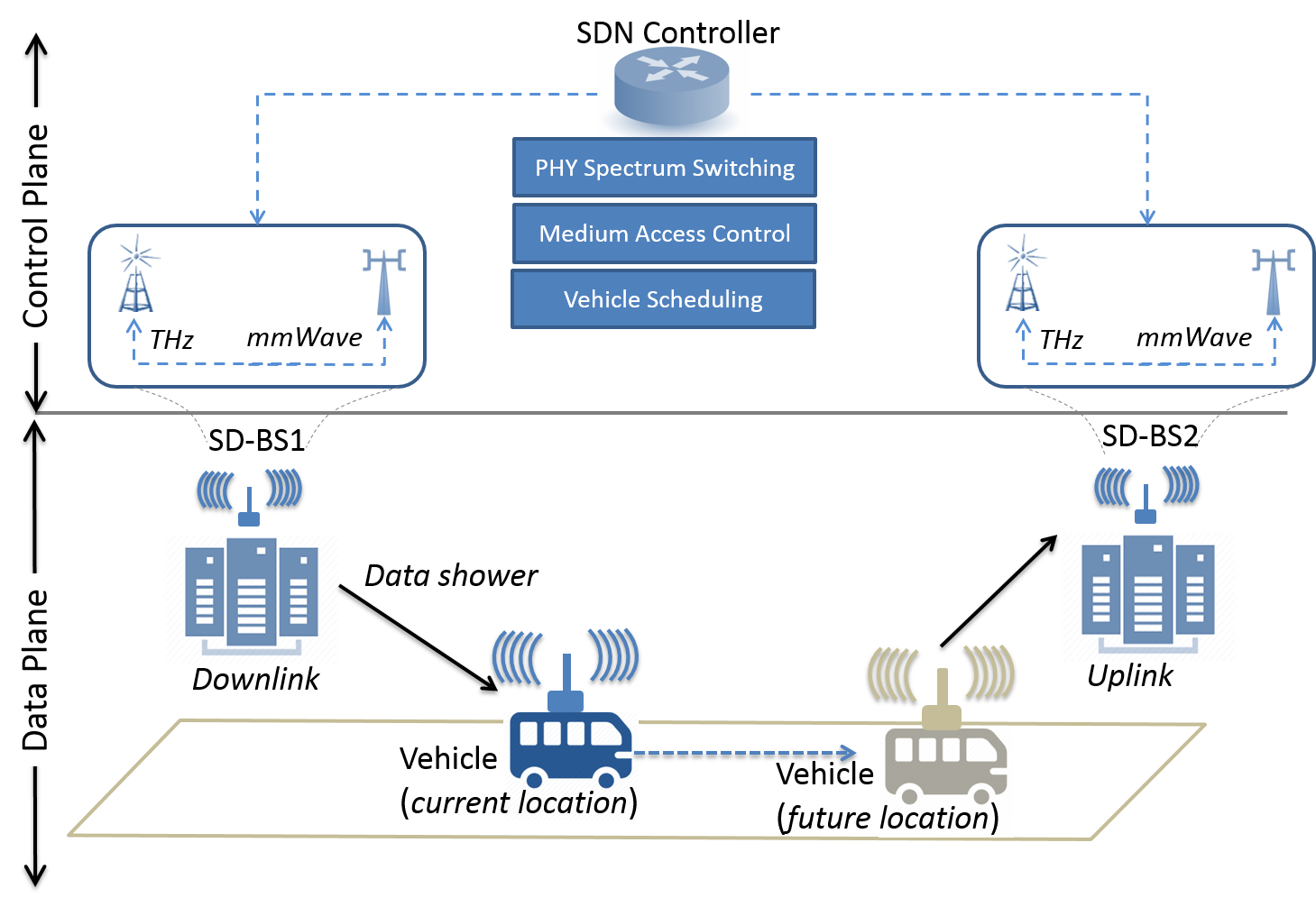}
\caption{Network architecture for SDN controlled mmWave/THz connections.}
\label{fig:arch}
\end{figure}

The \textit{infrastructure} in Fig.~\ref{fig:arch} refers to installed roadside software-defined base station (SD-BS) typical of small cells that operate under the directive of a SDN controller. Unlike the traditional cellular network, where base stations are spaced out in a hexagonal grid pattern, SD-BS are opportunistically placed and their locations can largely be random. Recent efforts have pointed towards the need of SDN-based resource sharing, by centralizing the physical and medium access control (MAC) functions, alongwith typical  operator tasks of load balancing and admission control policy. We adopt this approach in our work, wherein a SDN controller helps establish robust communication in mmWave frequencies where close to 7 GHz chunks of contiguous bandwidth are available. The mmWave band extends between $30-100\,\mathrm{GHz}$ (the upper boundary is loosely defined), and as a result, the wavelength of the transmitted signal lies between $1-10\,\mathrm{mm}$. The links so formed may be in the range of $200$ meters~\cite{RappTWC15}. This is considerably longer than the link distance in the THz bands, typically in the order of few meters. 

The 802.15 THz group report from March 2015 advocates even higher frequencies to `future-proof' the access technology, where frequencies in the 0.1-10 THz range could be used to achieve several Tbps transmission rates. However, this so called \textit{data shower} is possible only for very short distances of few meters. Common features in both these frequency ranges, however, are the practical need for directional transmission, accurate synchronization and occasional channel induced outages that may lead to bursty errors. When both types of wireless access become possible, there are several non-trivial tradeoffs that play a role in the SDN controller deciding which one of the two should be selected. The mmWave allows communication to commence at a greater separation distance, and thus can result in longer connected durations if there is relative motion between the nodes of the link. On the other hand, data exchange in the THz range may incur additional time for the node pair to be aligned in close proximity, but then it quickly ramps up by leveraging massive levels of bandwidth in such frequencies. There are additional considerations in this access selection problem, including the need for accommodating the channel-induced BER, which is unique for the choice of spectrum, and the amount of backlogged data to be delivered. 


As shown in Fig.~\ref{fig:arch}, vehicles connect to SD-BS 1 for very short access times (in the order of seconds) during their motion. Considering the example of data backhauling, they may download the desired data at that location, and then upload the data via target tower 2 when proximity conditions allow. By using vehicles as data mules, the source-destination SD-BSs themselves need not have direct LoS between their individual antennas or incur infrastructural deployment costs.

Our work on mmWave and THz-assisted data mule paradigm has the following contributions:
\begin{itemize}
\item \textbf{Dynamic THz/mmWave Spectrum Switching:} We design a new mode selection protocol that allows the SDN controller to decide when one of these (mmWave or THz) physical layers should be preferentially chosen for a given SD-BS to vehicle link, and develop handoff techniques between these two access technologies. 
\item \textbf{Capacity modeling:} We analytically derive effective data upload rates as a function of channel characteristics of mmWave and THz channels, SD-BS locations, and vehicular paths and obtain bounds on how much data can be delivered between two end points within a permissible time threshold. 
\item \textbf{Vehicle scheduling:} We propose an optimal admission policy at the SDN controller for scheduling multiple vehicles for accessing a given SD-BS, to account for practical road congestion scenarios, considering the heterogeneity of the mmWave and THz links. Since the search for the optimal scheduling is a NP-hard problem,  we design a computational-feasible greedy scheduling algorithm, exhibiting a polynomial-time complexity. 
\item \textbf{Simulation and case study:} We show the performance evaluation of our approach through simulations, as well as provide an example of a vehicle-assisted data backhauling considering the road topology of Boston city. 
\end{itemize}

The rest of this paper is organized as follows: Sec.~\ref{sec:2} describes the THz modeling background, related work and network assumptions. Sec.~\ref{sec:3} describes the THz/mmWave mode selection and medium access protocol while Sec.~\ref{sec:4} gives the analytic derivation of the capacity of our approach. Sec.~\ref{sec:4b} provides a method for scheduling the channel access for multiple vehicles. Sec.~\ref{sec:5} has a comprehensive performance evaluation, and conclusions are presented in Sec.~\ref{sec:6}.

\section{Background and Architectural Assumptions}
\label{sec:2}
In this section, we first describe the main propagation characteristics of the mmWave and the THz bands used in the next sections of the paper, and the underlying architectural assumptions.

\subsection{THz Channel Model}
\label{sec:2.1}
The signal propagation in the THz-band is mainly affected by molecular absorption, which results in both molecular absorption loss and molecular absorption noise \cite{JorAky11,JosGlob15, Han15}. In particular, the molecular absorption defines several transmission windows along the frequency scale with varying widths that are, to some extent, defined by the molecular composition of the medium. 

The THz channel transfer function $H_{\text{THz}}(f,d)$ consists of a spreading loss function and a molecular absorption loss function given by ~\cite{JorAky11,JorAky14}:
\begin{align}
\label{eq:THzchannel}
H_{\text{THz}}(f,d)=\frac{c}{4\pi f d} e^{-\frac{k(f) d}{2}} e^{-j 2 \pi f \tau_{\text{LOS}}},
\end{align}
where $c$ denotes the speed of light, $d$ stands for the distance between the transmitter and the receiver, and $\tau_{\text{LoS}} = d/c$ equals to the time-of-arrival of the line of sight (LoS) propagation. $k(f)$ is the frequency-dependent medium absorption coefficient that depends on the molecular composition of the transmission medium, i.e., the type and concentration of molecules found in the channel. Additional details for computing $k(f)$ and its effects on the THz propagation are reported in \cite{JorAky11}.
As in \cite{JorAky11,JorAky14}, in this paper we do not account for Non-Line-of-Sight (NLoS) transmissions in the THz band due to the lack of experimental characterization. The few NLoS channel models existing to date \cite{Han15} are mainly focused on the lower end of the THz band, i.e., 0.06 to 1 THz. We note that, by neglecting the NLoS opportunities, we underestimate the data shower in the THz band, i.e., we derive a lower bound on the achievable capacity in THz band. Moreover, we highlight that, by separately considering the NLoS propagation and outage event only for the mmWave communications, we incorporate in our model the fact that mmWave links are more robust than the THz links.

The molecular absorption determines not only the attenuation characteristics of the THz medium but also the noise. As described in \cite{JorAky11,JorAky14}, the noise can be modeled as additive, colored Gaussian. In our work, we denote the distance-dependent noise power spectral density (p.s.d.) as $S_{n}(f,d)$.
This model indicates that the THz channel is highly frequency-selective, and, in addition, the molecular absorption noise is non-white. Thus, the capacity can be obtained by dividing the total bandwidth $B_{\text{THz}}$ into many narrow sub-bands of width $\Delta f_i$ and summing the individual capacities \cite{JorAky11, Han15}. In fact, if the sub-band width is small enough, the channel appears as frequency-nonselective and the noise p.s.d. can be considered locally flat. Thus, by denoting with $N_{B}$ the number of sub-bands and with $f_i, i \in \{1,\ldots,N_B\}$ the center frequency of the $i$-th sub-band, the resulting capacity in bits/s is given by:
\begin{align}
\label{eq:cap_thz}
\mathcal{C}_{\text{THz}}^{\text{LOS}}(d)=\sum_{i=1}^{N_B} \Delta f_i \log{\left(1+\frac{|H_{\text{THz}}(f_i,d)|^2 P_i}{\Delta f_i S_{n}(f_i,d)}\right)}
\end{align}
where $P_i$ is the power associated to the $i$-th sub-band accounting for the antenna directional gains, under the constraint $\sum_{i=1}^{N_B} P_i \leq P_s$ with $P_s$ denoting the overall power, and
$H_{\text{THz}}(f,d)$ is reported in \eqref{eq:THzchannel}. From \eqref{eq:cap_thz}, as pointed out in \cite{JorAky11}, the THz channel capacity depends on the frequency $f_i$ of the electromagnetic wave, the transmission distance $d$,  the molecular composition of the channel through $H_{\text{THz}}(f,d)$ and $S_{n}(f,d)$, and the powers $P_i$.

\subsection{Architectural Assumptions}
\label{sec:2.2}
Software Defined Network (SDN)-based paradigm~\cite{AkyildizSDN} is needed for seamless communication brought about by efficient resource sharing, thereby achieving high spectral efficiency, when involving multiple different wireless technologies, namely, LTE, mmWave and THz, and providing support for mobility. The software definition enhances rapid prototyping and reconfiguring of protocols thereby allowing for flexible processing on the hardware at runtime. The SDN control plane implementing the centralized PHY/MAC functions enables the physical layer switching, running MAC layer chunk size determination algorithms and the medium access scheduling for multiple vehicles.

The network architecture involves the SDN controller, providing the necessary abstraction to applications, moving vehicle and, a SD-BS that have three different connectivity options: (i) classical LTE bands used only for control packets when data communication occurs in mmWave band, (ii) mmWave transceivers used for data primarily, but in a secondary role, for sending control packets when THz channels are used for data, and (iii) short distance THz transceivers that may be used for one directional data transfers only, at a given time. 

SD-BS can perform in-band signaling of real-time control messages, network status to the controller which in turn can feed back the control policies that best optimize for high link utilization~\cite{5GRoadmap} via standard interfaces like OpenFlow~\cite{OpenFlow}. Since OpenFlow is capable of providing a uniform interface for different wireless standards it enables user mobility when moving across SD-BSs that support multiple wireless standards.

In addition, the mules are equipped with caches able to fetch big amount of data. This is a very reasonable assumption since the available memory capacity is considered the fastest growing and yet untapped network resource today due to the continuous progress of the storage technology.

\noindent$\bullet$ \textit{Localization:} As the tower communication antennas are fixed and the vehicles today are generally equipped with GPS technology accurate to about a meter, we assume that there is full knowledge about the geolocation of both the mule and the tower antennas. Thus, the start/stop times for communication can be set accurately through beacons transmitted via currently existing and classical 802.11p/WAVE standards~\cite{80211p,Tsukamoto08}. Various techniques for tracking the sender/receiver during an ongoing communication have been proposed in THz channels, where a narrow-beam turns progressively thereby avoiding the need for frequent re-synchronization~\cite{JosGlob15}.~\cite{PrelcicRadar} uses out-of-band mmWave radar to aid beam alignment which significantly helps in reducing the beamsteering complexity. We account for the beamsteering complexity in the resulting overhead time. Finally, Doppler shift arising in the vehicular speeds of interest may also be discounted as directional and steerable antennas can mitigate the impact of relative motion~\cite{RappDopl14}. 

\noindent$\bullet$ \textit{Need for mode switching:} We incorporate the fact that mmWave links are more robust than the THz links by separately considering the non line of sight (NLoS) propagation and outage event for the former only. Thus, the THz links can be in two different states: a LoS path is available or there is an outage. For the mmWave links, recent work suggests that the states of LoS, NLoS and outage are distinct~\cite{RappTWC15}. Furthermore, experimental studies have demonstrated that the outage probability is small enough to be neglected, when the relative distance between the sender-receiver nodes is less than 200m~\cite{RappCh14}. Hence, when the distance between the mule and the tower is smaller than this threshold but greater than what is possible over THz link, the SDN controller prefers the mmWave link. Given the relative robustness of the mmWave link and high susceptibility of errors arising from NLoS in THz, the former can also be used as a separate control channel to return packet reception acknowledgments from the receiver to the sender that are communicating data in the THz channel. Our medium access protocol design assumes that both the mmWave and THz transceivers, albeit individually half-duplex and operating on entirely different frequency spectrum, can together be used to create a full duplex link.

\noindent$\bullet$ \textit{Noise-limited communications:} It is worthwhile to note that, given the highly directional nature of the  mmWave and THz  access technologies, the ensuing communication is not interference-limited; rather it is noise-limited. Hence, the concept of medium access protocol refers to the selection and configuration of the mmWave and THz communication modes, so that the maximum data transfer can be achieved along with the assurance of an error recovery capability. This is detailed discussed in the next section.

\section{Dynamic Spectrum Switching and Medium Access Protocol}
\label{sec:3}
The protocol described in this section is concerned with the selection and configuration of the mmWave and THz modes of communication at the SDN controller, so that (i) the maximum data transfer can be achieved, and (ii) error recovery can be assured.  

\subsection{Distance-dependent spectrum switching}
\label{sec:3_1}
\begin{figure}[!t]
\centering
\includegraphics[width=1\columnwidth]{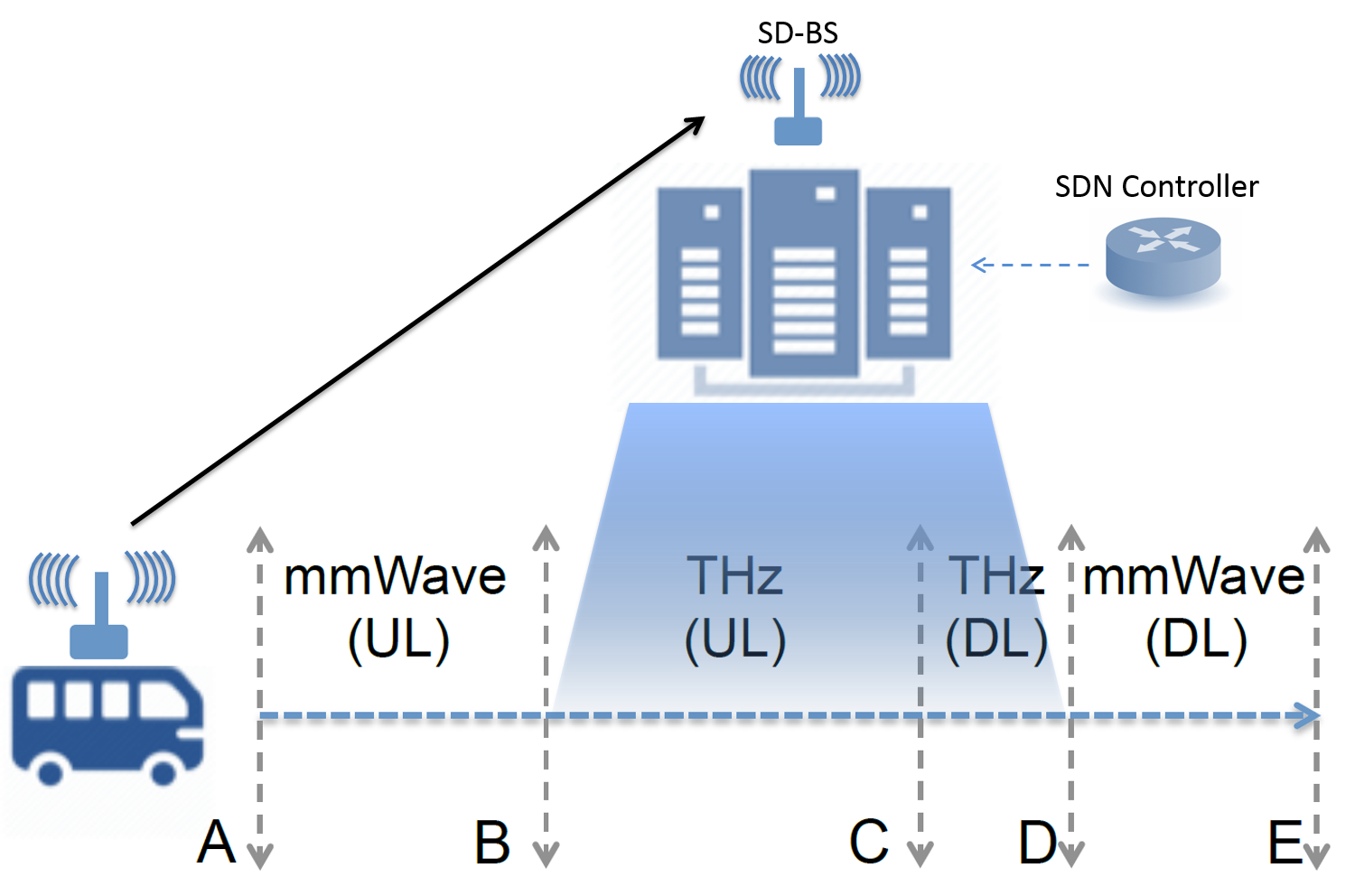}
\caption{Selecting durations for uplink (A-C) and downlink (C-E)}
\label{fig:updl}
\end{figure}

Let the maximum distances between a pair of nodes at which communication becomes possible for the mmWave and the THz channels be given by  $d_{\text{th}}^{\text{mm}}$ and $d_{\text{th}}^{\text{THz}}$, respectively. As discussed in Sec.~\ref{sec:2}, $d_{\text{th}}^{\text{mm}} >> d_{\text{th}}^{\text{THz}}$. As THz bands allow transmission rates of several orders of magnitude higher than mmWave, we propose to use this mode whenever possible. Thus, the communicating node pair {\em always} switches to THz communication when the separation distance is less than $d_{\text{th}}^{\text{THz}}$, and to mmWave band when $d_{\text{th}}^{\text{THz}} \leq d \leq d_{\text{th}}^{\text{mm}}$. For example, in Fig.~\ref{fig:updl}, the vehicle is moving from left to right, and in the process, reaching closer to the tower before pulling away again. The THz communication is only possible between B-D, and mmWave may be used both in A-B and D-E portions of the journey.

\subsection{Uplink/downlink optimization}
The overall data transfer between two physically separate towers requires downlink to the vehicle, the movement of the vehicle to the next location, followed by period of uplink. The vehicle repeats this cycle as it moves successively between the two infrastructure locations. As shown in Fig.~\ref{fig:updl}, we divide the interaction time of the vehicle with a tower into distinct uplink (UL) followed by downlink (DL) phases. The ratio of the time taken to complete these two phases is not fixed; rather it is negotiated on the classical LTE channel ahead of the vehicle's arrival in the vicinity of the tower. This depends upon the path  geometry-specific duration available for completing both the net UL/DL phases and the amount of backlogged data in either direction.

While a time-division like allocation for the UL and the DL phases allows us to clearly present the proposed framework, we note that the following analysis is neither affected by the directionality of the data shower, nor the assumption of time as a resource unit. Thus, our analytic derivations of capacity are valid for frequency-division or code-division resource allocations as well.

\subsection{Throughput maximization, packet aggregation, and error recovery}

\begin{figure}[!t]
\centering
\includegraphics[width=1\columnwidth]{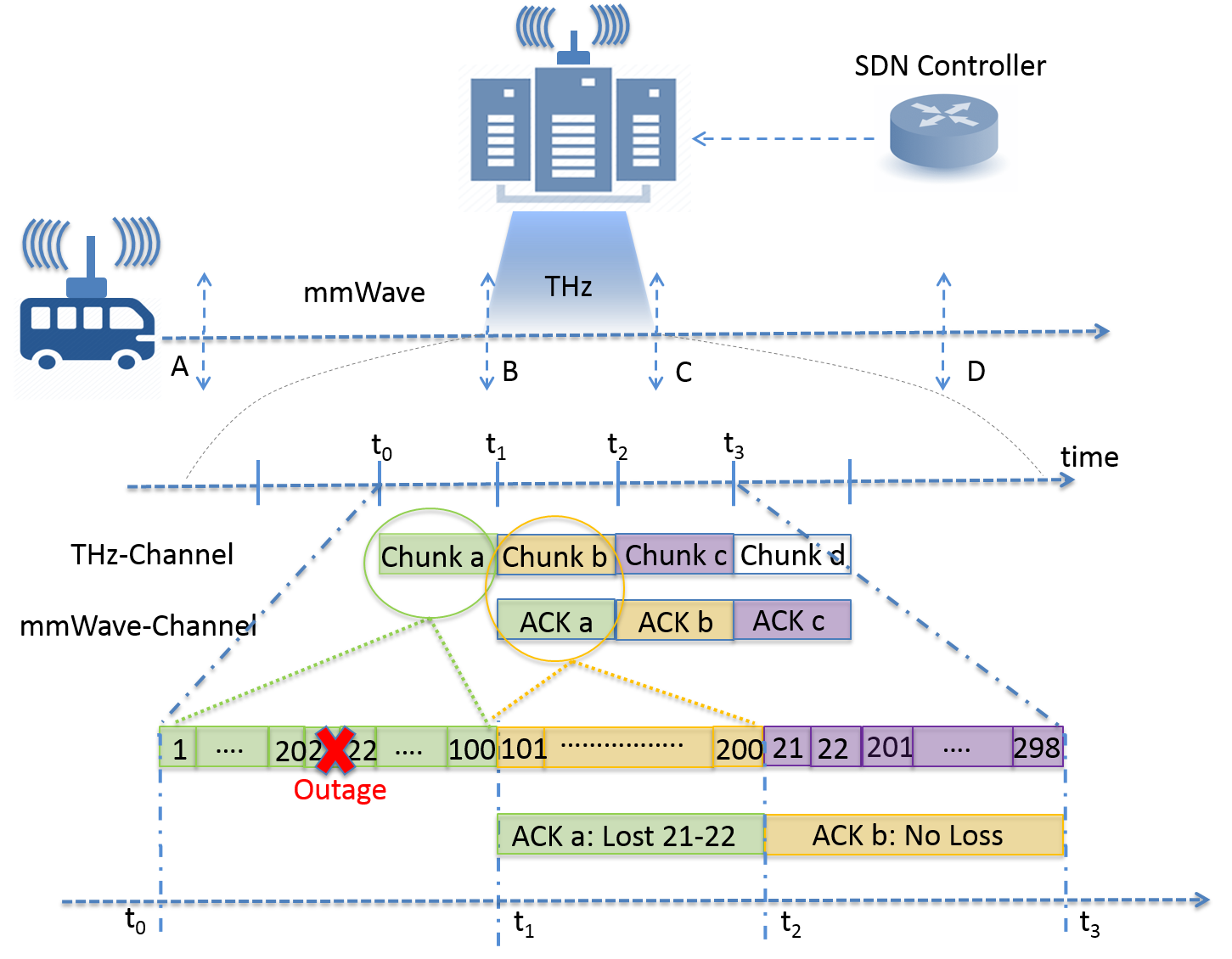}
\caption{Protocol overview for the uplink phase when the distance between  the mule and the tower antennas is smaller than the THz threshold. The data chunks are labeled with literals, whereas the numbers represent the packet IDs. A similar procedure applies when mmWave is used for data communication.}
\label{fig:3}
\end{figure}

According to the distance-dependent mode switching described in Section~\ref{sec:3_1}, the node pair always selects the best performing mode for data communication. To maximize the achievable throughput, we delegate the reverse path acknowledgments (ACKs) for the second-best option available at a given distance. Reversing the communication direction (for the ACK) introduces many challenges in completing a new round of beam training and alignment, synchronization etc. So instead, we retain the unidirectional flow of data in our scenario and delegate the slower and more reliable access technology for the returning ACKs. Specifically, when the THz link is active for one-way communication from sender to receiver, the mmWave link is used to report the ACKs from the receiver to the sender. When the mmWave is used for data communication, then the LTE link is used for ACKs reporting. Given that the transmission rate for data in each case is several order of magnitudes higher than that for the ACKs, the latter must be cumulative. We aggregate multiple data packets into a unit called as a \textit{data chunk}, and each ACK cumulatively validates the packets within the chunk. Our design saturates both the access technologies as ACKs are smaller, but for each mmWave ACK, there are at least an order of magnitude more data packets sent in the forward THz channel. The size of a chunk needs to be chosen so that both the forward (i.e., data) and the reverse (i.e., ACK) channels remain saturated. 
In summary, data packets are sent continuously without any gaps, and they are periodically validated with cumulative ACKs received through the reverse channel to allow efficient error recovery. In fact, when some packets of a data chunk are received with errors, these errors are notified back to the sender through second-best performing channel so that the sender can selectively re-transmit the lost data, but this time in the best-performing channel. As shown in Fig.~\ref{fig:3}, once the ACK is received through the reverse channel at the sender side, the lost packets are identified and re-transmitted within the next data chunk\footnote{Although the ACK processing delay could require that the lost packets will be re-transmitted at a some time slot in the future, we omit these particulars from Figure~\ref{fig:3} for the sake of simplicity.}, by prepending them to the new data. As a use-case, errors within the THz communication range are notified to the sender by using the mmWave band, allowing the sender to re-transmit in the active THz band. A similar process is used when ACKs are sent over LTE and data communication occurs over mmWave. In Fig.~\ref{fig:3}, two packets with IDs $21$ and $22$ belonging to the first chunk, say chunk $a$ transmitted at time $t_0$, are lost due to an outage event. The sender becomes aware of such a packet lost at time $t_2$, upon the reception of the corresponding ACK $a$. Hence, it re-transmits these two packets with the third chunk. Missing ACKs are handled in a conventional manner, i.e., the entire packet train (i.e., entire chunk) represented by that ACK will need to be re-sent in the forward channel. We assume constant chunk durations in this work, and propose to investigate the effect of dissimilar and derive optimal chunk intervals in future investigations.

\section{Capacity Modeling}
\label{sec:4}
In this section, we theoretically derive the effective channel capacity achievable through the proposed protocol, by exploiting both the mmWave and THz communications. In particular, the theoretical analysis accounts for the impact of relative distances and channel propagation conditions, noise and signal power.

\subsection{Capacity Formulation}
\label{sec:4.1}

We first introduce some definitions that will be used in the following analysis. Specifically, let us denote with $\mathcal{R}_{mm} \eqdef (d_{\text{th}}^{\text{THz}}, d_{\text{th}}^{\text{mm}}]$ the distance interval in which a mmWave communication is established. Similarly, we denote $\mathcal{R}_{THz} \eqdef (0,  d_{\text{th}}^{\text{THz}}]$ as the distance interval in which a THz communication is established.

\begin{defin}
\label{def_prob_mm}
$P(d)_{\text{mm}}^{\text{LoS}}$ denotes the probability of having a LoS connection between the transmitter and the receiver in the mmWave band, when their relative distance is $d$. $P(d)_{\text{mm}}^{\text{NLoS}}$ denotes instead the probability of having a NLoS connection between the transmitter and the receiver in the mmWave band, when their relative distance is $d$.
\end{defin}
Clearly such probabilities depend also on the geography of the considered network area, including building density and other natural/man made structures.

\begin{defin}
\label{def_prob_thz}
$P(d)_{\text{THz}}^{\text{LoS}}$ denotes the probability of having a LoS connection between the transmitter and the receiver in the THz band, when their relative distance is $d$.  
\end{defin}

We recall that the LTE interface is only used for ACKs and the mmWave link is used for data whenever the distance is $d_{\text{th}}^{\text{THz}} < d \leq d_{\text{th}}^{\text{mm}}$. Also, the mmWave interface is used for ACKs and the THz link for data whenever inter-node distance is $d \leq d_{\text{th}}^{\text{THz}}$. Hence, at a given relative distance $d$, the capacity $\mathcal{C}(d)$ available for transmitting data is given by:
\begin{align}
\nonumber
\mathcal{C}(d)&=\left[\mathcal{C}(d)_{\text{mm}}^{\text{LoS}} P(d)_{\text{mm}}^{\text{LoS}} + \mathcal{C}(d)_{\text{mm}}^{\text{NLoS}} P(d)_{\text{mm}}^{\text{NLoS}}\right] \mathrm{1}_{\mathcal{R}_{mm}}(d) +\\&
+ \left[ \mathcal{C}(d)_{\text{THz}}^{\text{LoS}}P(d)_{\text{THz}}^{\text{LoS}} \right] \mathrm{1}_{\mathcal{R}_{THz}}(d)
\label{eq:cap_dist}
\end{align}
where $P(d)_{\text{mm}}^{\text{LoS}}$, $P(d)_{\text{mm}}^{\text{NLoS}}$ and $P(d)_{\text{THz}}^{\text{LoS}}$ are defined in Definitions~\ref{def_prob_mm} and \ref{def_prob_thz}, respectively, and $\mathrm{1}_{\mathcal{R}_{mm}}(d)$ is the indicator function of $\mathcal{R}_{mm}$ given by:
\begin{equation}
\label{eq:ki}
\mathrm{1}_{\mathcal{R}_{mm}}(d)=
\begin{cases} 1, & d \in \mathcal{R}_{mm}
\\
0,  & \mathrm{otherwise.}
\end{cases}
\end{equation}
Similarly, $\mathrm{1}_{\mathcal{R}_{THz}}(d)$ is the indicator function of $\mathcal{R}_{THz}$. In \eqref{eq:cap_dist}, $\mathcal{C}(d)$ represents the available channel capacity at a given distance $d$, qualified further with appropriate subscripts (mm, THz) depending upon which access mode is used, and superscripts (LoS, NLoS) depending upon which of these propagation conditions exist. 

We stress that equation \eqref{eq:cap_dist} is valid regardless of the adopted models for the channel capacities and the probabilities of having LoS and NLoS paths. In the following, we expand \eqref{eq:cap_dist} further by considering some specific models for the channel capacity and the LoS and NLoS probabilities. 

\subsection{Case I - mmWave links}
First, regarding the mmWave capacity, when LoS link is available at a given relative distance $d$, we adopt the Shannon model used in \cite{RappCh14}: 
\begin{align}
\label{eq:cap_los}
\mathcal{C}(d)_{\text{mm}}^{\text{LoS}}=B_{\text{mm}} \log{\left(1+\gamma_{\text{mm}}(d)\right)},
\end{align}
where $\gamma_{\text{mm}}(d)$ denotes the average SNR, accounting for the directional antenna gains, observed at the distance $d$ in the mmWave spectrum of width $B_{\text{mm}}$. 

Second, when the NLoS link is available at a given relative distance $d$, we adopt the widely-used model that scales the LoS SNR with a factor $\Delta$ \cite{RappCh14}:
\begin{align}
\label{eq:cap_nlos}
\mathcal{C}(d)_{\text{mm}}^{\text{NLoS}}=B_{\text{mm}} \log{\left(1+\frac{\gamma_{\text{mm}}(d)}{\Delta}\right)}
\end{align}

Regarding the LoS and NLoS probabilities for mmWave communications, we adopt the models proposed in~\cite{RappCh14, RappTWC15}, since they were validated through experimental data. Specifically:
\begin{align}
\label{eq:p_los_mm}
P(d)_{\text{mm}}^{\text{LoS}}=\left(1-P(d)_{\text{mm}}^{\mathcal{O}}\right) e^{a_{\text{LoS}}d}
\end{align}
\begin{align}
\label{eq:p_nlos_mm}
P(d)_{\text{mm}}^{\text{NLoS}}=1-P(d)_{\text{mm}}^{\mathcal{O}}-P(d)_{\text{mm}}^{\text{LoS}}
\end{align}
where $P(d)_{\text{mm}}^{\mathcal{O}}$ denotes the outage probability that can be computed as~\cite{RappCh14, RappTWC15}:
\begin{align}
\label{eq:p_out_mm}
P(d)_{\text{mmW}}^{\mathcal{O}}=\max{\left(0, 1-e^{-a_{\mathcal{O}}d+b_{\mathcal{O}}}\right)},
\end{align}
In \eqref{eq:p_nlos_mm} and \eqref{eq:p_out_mm}, $a_{\text{LoS}}$, $a_{\mathcal{O}}$ and $b_{\mathcal{O}}$ are values empirically derived \cite{RappCh14, RappTWC15}.

\subsection{Case II - THz links}
Differently from mmWave communications, experimental data validating an outage distribution model is not available for the  THz band. Hence, by adopting a similar approach described in \cite{RappTWC15}, we assume an exponential distribution\footnote{This assumption is not restrictive, since the results derived within the paper continue to hold by simply adopting a different outage probability model.} for the outage in THz band as a function of the average SNR $\gamma_{\text{THz}}(d)$ at the distance $d$ \cite{Go05}:
\begin{align}
\label{eq:p_o_thz}
P(d)_{\text{THz}}^{\mathcal{O}}\eqdef 1-P(d)_{\text{THz}}^{\text{LOS}} =1-e^{-\gamma_{\text{th-THz}}/\gamma_{\text{THz}}(d)}
\end{align}
where $\gamma_{\text{th-THz}}$ denotes the minimum SNR required for establishing the THz link that depends also on the sensitivity of the receiver \cite{Go05}. 
Using the THz channel model described in Section~\ref{sec:2.1}, $\gamma_{\text{THz}}(d)$ can be evaluated as\footnote{In Section~\ref{sec:2.1} we assumed an ideal low-pass receiver filter. }:
\begin{align}
\nonumber
\gamma_{\text{THz}}(d)&=\frac{\int_{B_{\text{THz}}}S_t(f) |H_{\text{THz}}(f,d)|^2 df}{\int_{B_{\text{THz}}} S_n(f,d) df}=\\&
=\sum_{i=1}^{N_B} \frac{|H_{\text{THz}}(f_i,d)|^2 S_t(f_i)}{S_{n}(f_i,d)}
\label{eq:snr_thz}
\end{align}
Finally, regarding the THz capacity $\mathcal{C}(d)_{\text{THz}}^{\text{LoS}}$, it has been analyzed in Section~\ref{sec:2.1} and its expression is given in \eqref{eq:cap_thz}.   

\subsection{Data Shower Bulk}
Through the analysis described in the previous sections, we derived a closed-form expression for the channel capacity for a given distance $d$ when our network design is adopted. Using these results, we derive the maximum average number of data bits exchanged between the transmitter and the receiver in Proposition~\ref{prop_1}. We refer to this average number as \textit{data shower bulk}. We also provide in Corollary~\ref{cor_1} a closed-form expression for the \textit{data shower bulk} under the hypothesis of constant-speed straight trajectory. Before we proceed with this analysis, we list some preliminary definitions.

\begin{defin}
\label{def_2}
$\epsilon_{s}^{\text{mm}}$ denotes the time spent at the start of the mmWave communication to synchronize the transmitter and the receiver. This time is needed to calibrate the transceivers at a finer granular level, as observed in Section~\ref{sec:2}, despite the assumption of steerable antennas. Similarly, $\epsilon_{s}^{\text{THz}}$ denotes the time spent at the start of the THz communication to synchronize the transmitter and the receiver at a finer granular level. Finally, $\epsilon_{tr}$ denotes the time spent in switching from transmitting mode to receiving mode and vice versa.
\end{defin}

The time $\epsilon_{tr}$ takes into account not only the effective time for mode selection, but also a guard time to handle possible burst errors arising from the previous phase. The setting of such a parameter is beyond the scope of this paper, but does raise interesting design possibilities that we intend to explore in our future work. We observe that the optimization of such a parameter should account for the allocation strategy chosen for the UL and the DL phase, as well as the length of the  packet chunks which in turns depend on both the channel conditions and the delay propagation.

\begin{defin}
\label{def_3}
$t_{\text{in}}$ and $t_{\text{out}}$ denote the starting and the ending time of a contact event, respectively, i.e., the first and the last time instant in which the transmitter and the receiver could establish and sustain either a mmWave or THz link in a one-way journey.
\end{defin}

\begin{prop}
\label{prop_1}
The data shower bulk transferred by adopting the proposed architecture is given by:
\begin{align}
\label{eq:bit_tot_ex}
\overline{n}= \displaystyle\int_{t_{\text{in}}}^{t_{\text{out}}} \displaystyle\mathcal{C}(d(t)) dt
\end{align}
where $d(t)$ denotes the transmitter-receiver relative distance at time $t$ and the capacity $\mathcal{C}(d(t))$ is given in \eqref{eq:cap_dist}.  
\begin{IEEEproof}
See Appendix~\ref{app_1}
\end{IEEEproof}
\end{prop}

\begin{rem}
We note that the time interval $[t_{\text{in}}, t_{\text{out}}]$ can be characterized by a sequence of time-separated contact periods, as a consequence of the mule moving repetitively in and out of the communication range due to the street topology constraints. Nevertheless, the time instants belonging to the considered time interval at which the distances $\{d(t)\}$ do not range in $\{\mathcal{R}_{mm} \bigcup \mathcal{R}_{THz}\} $ do not contribute to the transferred bits $\overline{n}$, since the capacity $\mathcal{C}(d(t))$ is null according to \eqref{eq:cap_dist}. 
\end{rem}

\begin{rem}
The data shower bulk $\overline{n}$ derived in Proposition~\ref{prop_1} constitutes an upper bound of the layer-2 throughput achievable by adopting the proposed architecture. In fact, \eqref{eq:bit_tot_ex} does not account for the synchronization overhead associated with the times $\epsilon_{s}^{\text{mm}}$ and $\epsilon_{s}^{\text{THz}}$, as well as the switching overhead associated with the time $\epsilon_{tr}^{\text{THz}}$. Furthermore, the throughput depends on a number of physical-realization parameters, such as the adopted channel code, the adopted modulation technique, as well as the synchronization techniques and the mode switching procedure.
\end{rem}

In the following we derive in Corollary~\ref{cor_1} a strict bound for the data shower bulk, under the hypothesis of uniform straight movement from A to E as depicted in Fig.~\ref{fig:updl}. To this aim, let us denote with the $\alpha$ the angle formed by: (i) the distance $d_{\text{th}}^{\text{mm}}$ between the mule with the tower at time $t_{\text{in}}$, and (ii) the direction of the movement.

\begin{cor}
\label{cor_1}
The data shower bulk transferred by adopting the proposed architecture under the hypothesis of constant-speed straight-trajectory with average speed $v$ is given by:
\begin{align}
\label{eq:bit_tot}
\overline{n}=\left( \frac{2 d_{\text{th}}^{\text{mm}} \cos \alpha}{v}-\epsilon_{s}^{\text{mm}}-\epsilon_{s}^{\text{THz}}-\epsilon_{tr}\right) \frac{\displaystyle\int_{d_{\text{min}}}^{d_{\text{th}}^{\text{mm}}}\displaystyle\mathcal{C}(\eta) d\eta}{d_{\text{th}}^{\text{mm}}-d_{\text{min}}}
\end{align}
where $d_{\text{min}}$ is the minimum distance between the antennas of the mule and the BS during the movement, $\mathcal{C}(d)$ is given in \eqref{eq:cap_dist} and $\epsilon_{s}^{\text{mm}}, \,\epsilon_{s}^{\text{THz}}, \,\epsilon_{tr}^{\text{THz}}$ are defined in Definition~\ref{def_2}.  
\begin{IEEEproof}
See Appendix~\ref{app_2}
\end{IEEEproof}
\end{cor}

The data shower bulk $\overline{n}$ derived in Corollary~\ref{cor_1} constitutes a stricter bound than \eqref{eq:bit_tot_ex}. In fact, in \eqref{eq:bit_tot} we explicit some time overhead through $\epsilon_{s}^{\text{mm}}, \,\epsilon_{s}^{\text{THz}}, \,\epsilon_{tr}^{\text{THz}}$.

\section{Multi-vehicle Scheduling}
\label{sec:4b}
The work so far covers the capacity formulation for a single vehicle exchanging data with one roadside infrastructure location. However, multiple vehicles $\mathcal{V} = \{1, \ldots, V\}$ may also pass through the same region concurrently. This requires the SDN controller scheduling them at different time instants (there is only one mmWave/THz transceiver at the roadside location) so that all their cumulative bandwidth needs are satisfied. The scheduling time is dependent also on the location of the vehicles at that instant, which in turn influences whether the mmWave or the THz link is active. 

Considering that the entire time horizon is composed of slots of duration $T$, and let $D_v$ with $v \in \mathcal{V}$ denote the number of bits uploaded/downloaded to/from the infrastructure tower that is bounded by $\bar{n}$ derived earlier in (\ref{eq:bit_tot_ex}). Further, let the number of  vehicles in $\mathcal{V}$ that are close enough to the tower so that a communication (either mmWave or THz) link can be established in a given time slot $k$ be given as $\mathcal{V}^k \subseteq \mathcal{V}$. Thus,
\begin{equation}
\label{eq:x.1}
\mathcal{V}^k = \{ v \in \mathcal{V} : d^k_v \in \mathcal{R}_{\text{mm}} \cup \mathcal{R}_{\text{THz}}\}
\end{equation}
with $d^k_v$ denoting the maximum distance of the $v$-th vehicle from the tower during time slot $k$, and $\mathcal{R}_{\text{mm}}$ and $\mathcal{R}_{\text{THz}}$ defined in Sec.~\ref{sec:4.1}. Note this implies that a vehicle belongs to $\mathcal{V}^k$ if and only if its distances in the entire time slot belong to $\mathcal{R}_{\text{mm}} \cup \mathcal{R}_{\text{THz}}$. 
In the following, for the sake of simplicity and without loss of generality, we assume $\underset{k}{\cup} \mathcal{V}^k = \mathcal{V}$.


\subsection{Scheduling Problem Formulation}
We devise an optimization problem to select which slots must be assigned to each vehicle $v \in \mathcal{V}$ with the objective of maximizing the total number of bits exchanged between the infrastructure tower and the vehicles within the time horizon under the following constraints:
\begin{itemize}
	\item[i)] the total number of bits exchanged with the $v$-th vehicle within a one-way journey does not exceed $D_v$;
	\item[ii)] at most one vehicle is scheduled during each time slot.
\end{itemize}

Constraint (i) avoids sub-optimal scheduling, i.e., it avoids assigning a time slot to a vehicle that already completed its communication needs (represented by $D_v$). Constraint (ii) accounts for the THz/mmWave mode selection described in Sec.~\ref{sec:3}. Although two technologies (mmWave/THz) can be concurrently used in a given time slot, only one vehicle can be scheduled in each time slot, since we exploit the second-best technology for reverse path acknowledgments.

By denoting with $N$ the total number of exchanged bits and with $N_v$, $v \in \mathcal{V}$, the number of bits exchanged with the $v$-th vehicle within a one-way journey, we can reformulate the considered problem as follows: 
\begin{align}
\label{eq:x.2}
& \text{given } k_s \leq k_e : & & \mathcal{V}^{k_s - 1} = \mathcal{V}^{k_e + 1} = \emptyset \\
\label{eq:x.3}
& & & \mathcal{V}^{k} \neq \emptyset , \forall \, k \in \mathcal{K} \\
\label{eq:x.4}
& \underset{\displaystyle\{\phi^k_v\}_{\substack{k \in \mathcal{K},\\v \in \mathcal{V}}}}{\text{maximize}} & & N \\
\label{eq:x.5}
& \text{subject to} & & N_v \leq D_v , \forall \, v \in \mathcal{V} \\
\label{eq:x.6}
& & & \sum_{v \in \mathcal{V}^k} \phi^k_v = 1 , \forall \, k \in \mathcal{K}
\end{align}
with $\mathcal{K} \eqdef \{k_s, \ldots, k_e\}$ denoting the set of time slots and
$\phi^k_v$ denoting the indicator function mapping each vehicle with a time slot, i.e., $\phi^k_v = 1$ if the $v$-th vehicle is scheduled within the $k-$th time slot and $\phi^k_v = 0$ otherwise. 

\eqref{eq:x.3} guarantees that, during each time slot of the considered time horizon, there exists at least one vehicle in connection with the tower. In fact, an \textit{empty} time slot represents a separation between different journeys, which need to be individually optimized due to the finite cache sizes. This is accounted for in \eqref{eq:x.2}. 

We derive in Proposition~\ref{prop:x.1}, the closed-form expression of $N$ and $N_v$, by accounting for the time spent by the tower and the vehicle to establish a physical link. This time depends on vehicle-specific parameters, such as the adopted antenna beamforming algorithm. To abstract the scheduling problem from underlying dependencies, we accumulate all such coordination overheads within $T^o_v > 0$, which we refer to as \textit{overhead time}.

\begin{algorithm}[t]
\caption{Greedy Scheduling Algorithm}
\label{alg:x.1}
 {\small
\begin{algorithmic}[1]
\STATE $\phi^k_v = 0, \forall \, k \in \mathcal{K} \wedge v \in \mathcal{V}$
\FORALL{$k \in \mathcal{K}$} \label{line:x.1.s}
	\IF{$| \mathcal{V}^k| == 1$}
        \STATE \COMMENT Only 1 vehicle in contact
		\STATE $v = \mathcal{V}^k[1]$\\
		\COMMENT with $A[n]$ denoting the $n$-th element of array $A$
		\STATE $\phi^k_v = 1$
		\STATE $\mathcal{K} = \mathcal{K} \setminus \{k\}$ \label{line:x.9}
		\STATE $D_v = D_v - \tilde{n}^k_v$ \label{line:x.2.s}
		\IF{$D_v \leq 0$}
			\STATE $\mathcal{V}^k = \mathcal{V}^k \setminus \{ v \} \; \forall \, k \in \mathcal{K}$ \label{line:x.3} \\
		\ENDIF \label{line:x.2.e}
	\ENDIF
\ENDFOR \label{line:x.1.e}
\FORALL{$k \in \mathcal{K}$} \label{line:x.4.s}
	\IF{$\mathcal{V}^k == \emptyset$}
		\STATE $\mathcal{K} = \mathcal{K} \setminus \{k\}$
        \STATE \COMMENT Remove empty slot
	\ENDIF
\ENDFOR \label{line:x.4.e}
\WHILE{$\mathcal{K} \neq \emptyset$} \label{line:x.5.s}
	\STATE $k_t,v_t \leftarrow$ Algorithm~\ref{alg:x.2} \label{line:x.6}
	\STATE $\phi^{k_t}_{v_t} = 1$
	\STATE $D_{v_t} = D_{v_t} - \tilde{n}^{k_t}_{v_t}$
	\STATE $\mathcal{K} = \mathcal{K} \setminus \{k_t\}$ \label{line:x.10}
	\IF{$D_{v_t} \leq 0$}
		\FORALL{$k \in \mathcal{K} : v \in \mathcal{V}^{k}$}
			\STATE $\mathcal{V}^k = \mathcal{V}^k \setminus \{ v_t \}$
			\IF{$\mathcal{V}^k == \emptyset$}
				\STATE $\mathcal{K} = \mathcal{K} \setminus \{k\}$
			\ENDIF
		\ENDFOR
	\ENDIF
\ENDWHILE \label{line:x.5.e}
\end{algorithmic}
}
\end{algorithm}

\begin{prop}
\label{prop:x.1}
The total number of bits $N$ exchanged between the tower and the vehicles is equal to
\begin{equation}
\label{eq:x.7}
N = \sum_{\substack{ v \in \mathcal{V}}} N_v
\end{equation}
where
\begin{equation}
\label{eq:x.8}
N_v = \displaystyle \sum_{k \in \mathcal{K}} \phi^k_v n_v^k
\end{equation}
and
\begin{equation}
\label{eq:x.9}
n_v^k=\displaystyle \int_{(k-1) T + \chi^k_v T^O_v}^{kT} C_v(d_v(t)) dt
\end{equation}
\begin{equation}
\label{eq:x.10}
\chi^k_v = \begin{cases}
		1 & \text{if } \phi^k_v - \phi^{k-1}_v = 1\\
        0 & \text{otherwise}
	\end{cases}
\end{equation}
with $C_v(d_v(t))$ given in (\ref{eq:cap_dist}) and $\phi^0_v \eqdef 0$.
\begin{IEEEproof}
See Appendix~\ref{app_prop_scheduling}.
\end{IEEEproof}
\end{prop}

\begin{rem}
\label{rem:x.1}
The scheduling problem is NP-hard, since: i) the variables $\phi^k_v$ denoting the scheduling-state of the $v$-th vehicle at time slot $k$ have integer values (actually, binary); ii) the presence of the overhead time $T_O$ within the integral in \eqref{eq:x.9}. In fact, the time complexity grows with the total number of possible solutions, i.e., $\mathcal{O}\left(V^K\right)$, where $V=|\mathcal{V}|$ is the number of vehicles and $K=|\mathcal{K}|$ is the number of time slots. 
As an example, when $V = 4$ and $K= 20$, it results $V^{K} = 4^{20} \simeq 10^{12}$. Hence, we design a greedy scheduling algorithm (see Algorithm~\ref{alg:x.1}), which has polynomial-time complexity.
\end{rem}


\begin{algorithm}[t]
\caption{Inner Procedure}
\label{alg:x.2}
 {\small
\begin{algorithmic}[1]
\STATE $\mathcal{K}_t = \argmax{k \in \mathcal{K}}{ \left\{ \displaystyle \max_{v \in \mathcal{V}^k} \{\tilde{n}^k_v\} \right\} }$\\
\IF{$|\mathcal{K}_t|$ == 1} 
	\STATE $k_t = \mathcal{K}_t[1]$
    \STATE $v_t = \argmax{v \in \mathcal{V}^{k_t}}{\{\tilde{n}^{k_t}_{v}\}}$\\
	\COMMENT if argmax returns multiple items, pick one at random
\ELSE \label{line:x.7.s}
	\FORALL{$i=1:|\mathcal{K}_t|$}
		\STATE $\mathcal{V}_t[i] = \argmax{v \in \mathcal{V}^{\mathcal{K}_t[i]}}{\{\tilde{n}^{\mathcal{K}_t[i]}_{v}\}}$\\
\ENDFOR
	\STATE $i_t = \argmin{i = 1, \ldots, |\mathcal{K}_t|}{ \left\{ \displaystyle \sum_{v \in \mathcal{V}^{\mathcal{K}_t[i]} \setminus \{\mathcal{V}_t[i]\} } \{\tilde{n}^{\mathcal{K}_t[i]}_v\} \right\} }$ \label{line:x.8} \\
	\STATE $k_t = \mathcal{K}_t[i_t]$
	\STATE $v_t = \mathcal{V}_t[i_t]$
\ENDIF \label{line:x.7.e}
\RETURN $k_t,v_t$
\end{algorithmic}
}
\end{algorithm}

\subsection{Explanation of Algorithm~\ref{alg:x.1}}

The greedy algorithm works by first (lines~\ref{line:x.1.s}-\ref{line:x.1.e}) computing the sets of slots during which only one vehicle can establish a communication link with the tower. For each such slot, a given vehicle is scheduled if the constraint described in equation \eqref{eq:x.5} is satisfied (lines~\ref{line:x.2.s}-\ref{line:x.2.e}). As soon as a vehicle completes its communication needs, it is excluded (line~\ref{line:x.3}) from all the remaining slots. As a consequence, a slot may become \textit{empty}, i.e. the number of vehicles to be scheduled could become zero. Lines-\ref{line:x.4.s}-\ref{line:x.4.e} remove such empty slots from $\mathcal{K}$.

In lines~\ref{line:x.5.s}-\ref{line:x.5.e}, the algorithm schedules vehicle $v_t$ at time slot $k_t$, if this choice maximizes the number of exchangeable bits without accounting for the scheduling overhead, i.e.:
\begin{equation}
\phi^{k_t}_{v_t} = 1 \Longleftrightarrow  \tilde{n}^{k_t}_{v_t} = \max_{k \in \mathcal{K} } \left\{  \max_{v \in \mathcal{V}^k} \{ \tilde{n}^k_v \} \right\}
\end{equation}
where
\begin{equation}
\tilde{n}^k_v = \int_{(k-1) T}^{kT} C_v(d_v(t)) dt
\end{equation}
Specifically, at line~\ref{line:x.6} $k_t$ and $v_t$ are computed through Algorithm~\ref{alg:x.2}, and the remaining lines schedule the vehicle and satisfy the constraint given in \eqref{eq:x.5}.

We note that Algorithm~\ref{alg:x.2}, through lines~\ref{line:x.7.s}-\ref{line:x.7.e}, accounts for the case in which multiple feasible choices for maximizing the number $\tilde{n}^k_v$ of transferred bits is possible, i.e., there exists multiple time slots in which the same maximum $\tilde{n}^{k_t}_{v_t}$ is achieved. In such a case, line~\ref{line:x.8}
selects the time slot in which the lowest communication opportunities (i.e., the lowest average number of exchangeable bits) are available to the remaining vehicles.

Finally, we note that the constraint given in equation \eqref{eq:x.6} is satisfied with lines~\ref{line:x.9} and \ref{line:x.10}. Hence, Algorithm~\ref{alg:x.1} computes a valid (admissible) solution for the considered scheduling problem.

\begin{rem}
The time complexity of the greedy algorithm given in Algorithm~\ref{alg:x.1} is $\mathcal{O}(V \cdot K^2)$. Specifically,  Algorithm~\ref{alg:x.1} exhibits a polynomial complexity, which grows quadratically with the number of time slots and linearly with the number of vehicles. Clearly, this is an attractive feature since it assures the computational practicability of the algorithm. With reference to the example given in Remark~\ref{rem:x.1}, it results $V \cdot K^2 \simeq 2 \cdot 10^3 \ll 10^{12}$.
\end{rem}

\section{Data Exchange Evaluation}
\label{sec:5}

\begin{figure}[t]
\centering
\includegraphics[width=1\columnwidth]{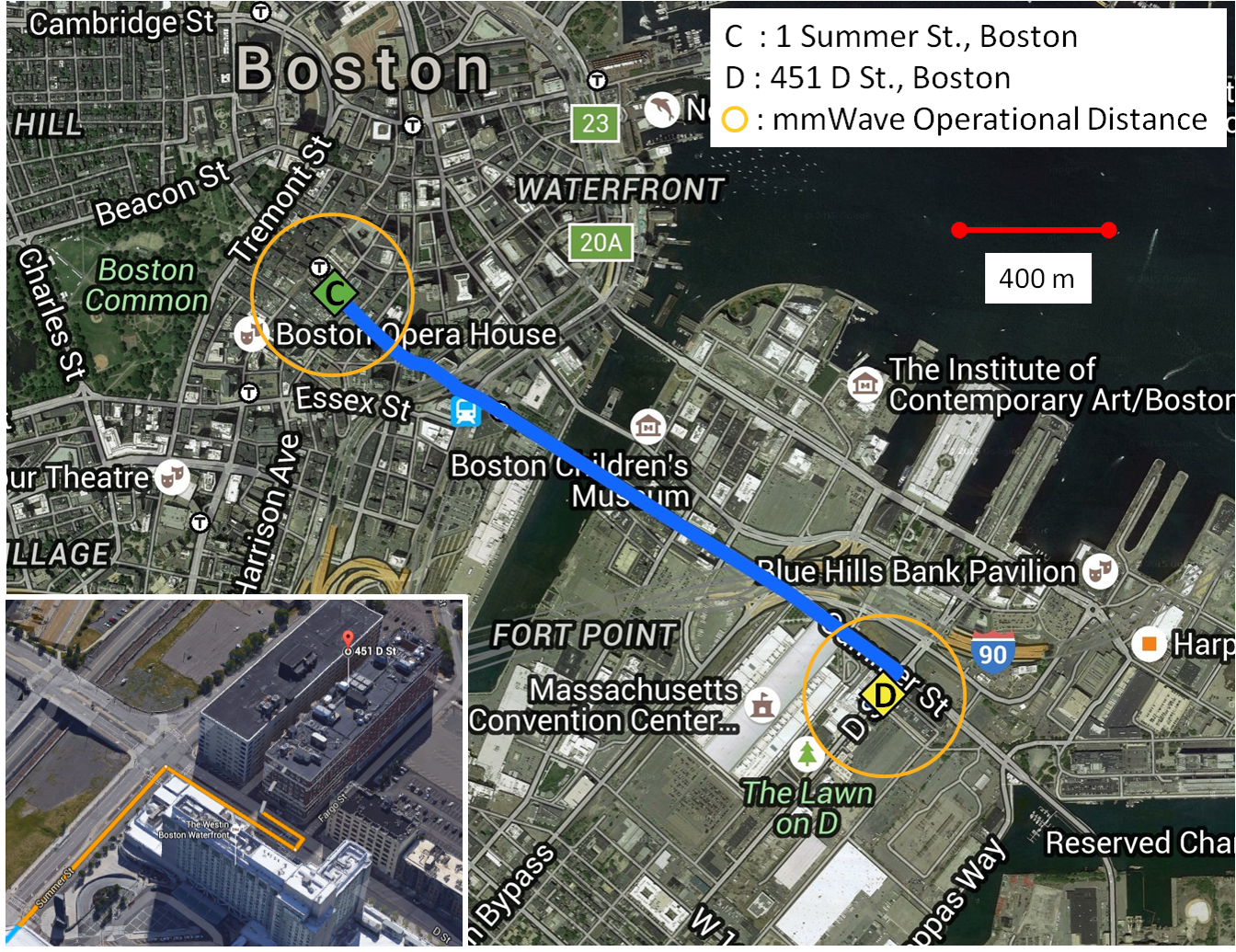}
\caption{Google maps showing the suggested route for a vehicle moving from \textit{1 Summer Street} to \textit{451 D St.}. The end-to-end distance is roughly 1.2 miles and the estimated travel time is about 7 minutes, depending on the traffic conditions. The yellow circle represents the mmWave operational distance.}
\label{fig:b}
\end{figure}


\begin{table}
\centering
\caption{Parameter Setting}
\begin{tabular}{l c}
mmWave parameter & Value\\
\hline
$f_c$: carrier frequency & $73$ GHz \\
$\Delta f_c$: uplink/downlink shared bandwidth & $1$ GHz\\
$\alpha$: path-loss intercept least squares fit  & LoS: 69.8, NLoS: 82.7\\
$\beta$: path-loss slope least squares fit & LoS: $2$ - NLoS: $2.69$\\
$P_{tx}$: transmit power & $30$ dBm\\
$G$: directional antenna gain & $27$ dB\\
Noise power & -$87$ dBm\\
Noise figure & $5$ dB\\
$d_{\text{th}}^{\text{mm}}$: operational distance & $200$ m\\
$1/a_{\text{LoS}}$: LoS state probability parameter & $37$ m\\
$1/a_{\text{O}}$: outage state probability parameter & $45.5$ m\\
$1/b_{\text{O}}$: outage state probability parameter & $3.3$\\
\hline\\
THz parameter & Value\\
\hline
$k(f)$: frequency-dependent coefficient & $[2 \cdot 10^{-6}-3 \cdot 10^1]cm^{-1}$\\
$f_c$: carrier frequency & $0.85$ THz \\
$\Delta f_c$: uplink/downlink shared bandwidth & $0.1$ THz\\
$P_{tx}$: transmit power & $0 - 20$ dBm\\
$G$: directional antenna gain & $27$ dB\\
$d_{\text{th}}^{\text{THz}}$: operational distance & $10$ m\\
\hline
\end{tabular}
\label{tab:1}
\end{table}

\begin{figure*}[t]
\begin{minipage}[c]{0.48\textwidth}
\includegraphics[width=1\columnwidth]{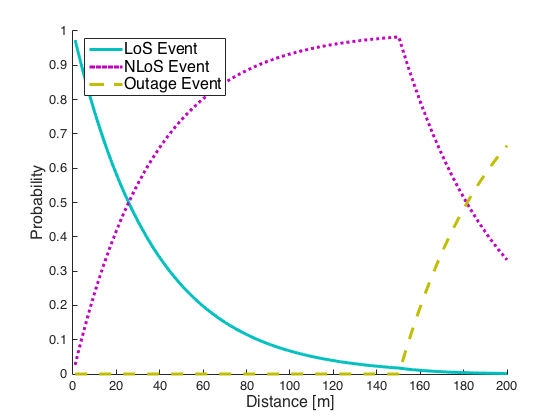}
\caption{Empirical LoS, NLoS and outage probabilities for a mmWave link at $73$GHz as a function of the separation distance between transmitter and receiver.}
\label{fig:f}
\end{minipage}
\hspace{0.02\textwidth}
\begin{minipage}[c]{0.48\textwidth}
\includegraphics[width=1\columnwidth]{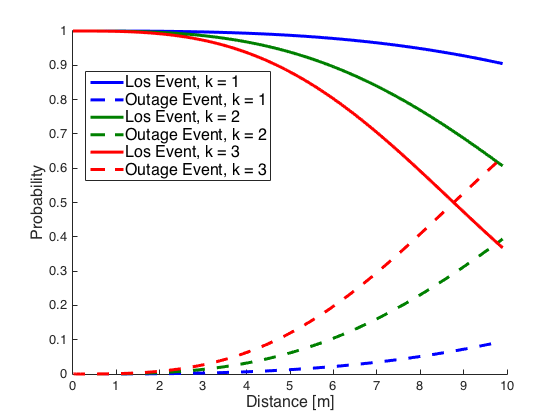}
\caption{LoS and Outage probabilities for a THz link at $0.85$THz with $0$dBm transmitted power as a function of the separation distance between transmitter and receiver.}
\label{fig:g}
\end{minipage}
\end{figure*}


\begin{figure*}[t]
\centering
\begin{minipage}[c]{0.48\textwidth}
\includegraphics[width=1\columnwidth]{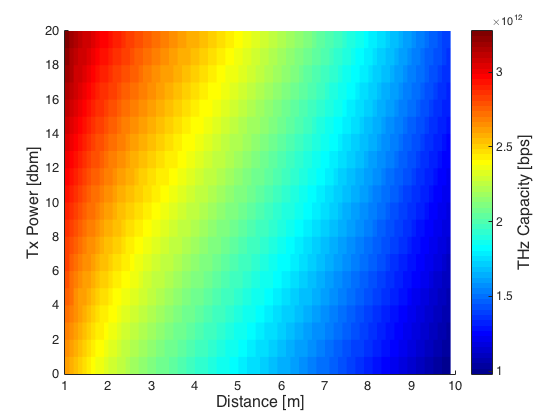}
\caption{Capacity for a THz link at $0.85$THz  as a function of the transmitted power and the separation distance between transmitter and receiver. Outage events have been considered.}
\label{fig:h}
\end{minipage}
\hspace{0.02\textwidth}
\begin{minipage}[c]{0.48\textwidth}
\includegraphics[width=1\columnwidth]{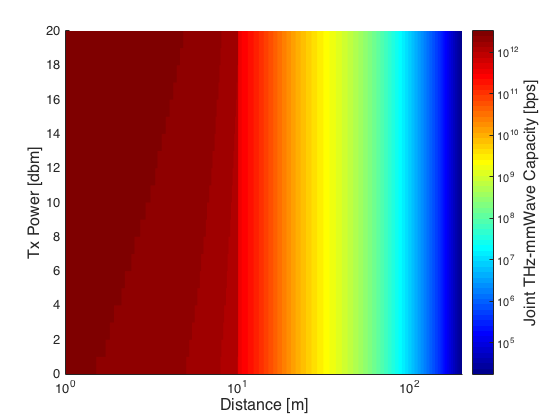}
\caption{Capacity achievable by adopting the proposed THz/mmWave mode selection, as a function of the transmitted power in the THz band and the separation distance between transmitter and receiver.}
\label{fig:i}
\end{minipage}
\end{figure*}

In this section, we evaluate the achievable capacity using an example scenario of V2I communication enabling data center traffic backhauling~\cite{Mahimkar11}.

Specifically, we first introduce the adopted scenario in Section~\ref{sec:5.1}. Then,we assess the capacity as a function of the distance for both the mmWave and the THz links in Section~\ref{sec:5.2}. In Section~\ref{sec:5.3}, we derive the data shower bulk as a function of the minimum distance between the transmitter and receiver antennas, along with the effective data transfer rates for data centers located in Boston. Finally, in Section~\ref{sec:5.4}, we assess the benefits of adopting the proposed multiple-vehicle scheduling algorithm.

\subsection{Network Scenario}
\label{sec:5.1}
To assess the achievable amount of exchanged data for backhauling under realistic conditions, we consider the actual positions of existing data centers located in Boston city \cite{map}.

Out of 22 available data centers, we choose two centers located in downtown Boston as typical use case: the first is located at \textit{1 Summer Street}, owned by \textit{XO Communications}, and the second is located at \textit{451 D St.}, owned by \textit{Markley Group LLC}. Through Google Maps, we obtain the suggested vehicular route between the two considered data centers, shown in Fig.~\ref{fig:b}. The vehicle route length is roughly $1.2$ miles long with an estimated travel time ranging between 7 and 19 minutes. The inline picture shows the zoomed in view of the route near the first center. This is to indicate that the journey does account for the constraints arising from buildings and lanes.

From Google maps directions, basing on the antenna positions, we can estimate the distance between transmitter and receiver as function of time. We emulate a vehicle-assisted deployment where antennas are placed on vehicle rooftops and streetlight poles closest to the chosen data center, respectively. The rationale for this choice is twofold: i) the corresponding antenna heights agree with those used in mmWave channel measurements \cite{RappCh14} allowing so us to adopt the corresponding experimental mmWave channel model; ii) the antenna positioning ensure that the THz link is not affected by outage events caused by pedestrians or vehicles blocking the LoS path.

\begin{figure*}[t]
\centering
\begin{minipage}[c]{0.48\textwidth}
\includegraphics[width=1\columnwidth]{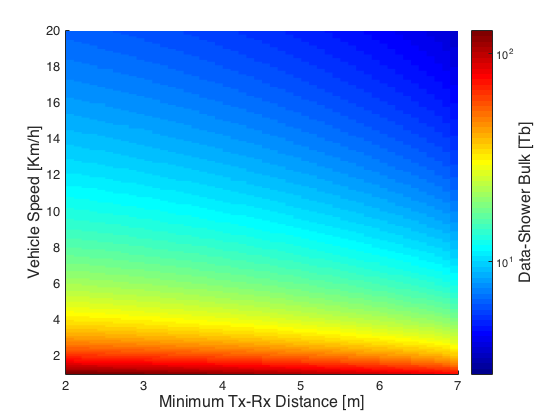}
\caption{Data Shower Bulk as a function of the minimum separation distance $d_{\text{min}}$ between the transmitter and the receiver and the average mule velocity. Single-way journey between the vehicle, moving with constant-speed along a straight-trajectory, and the tower.}
\label{fig:l}
\end{minipage}
\hspace{0.02\textwidth}
\begin{minipage}[c]{0.48\textwidth}
\includegraphics[width=1\columnwidth]{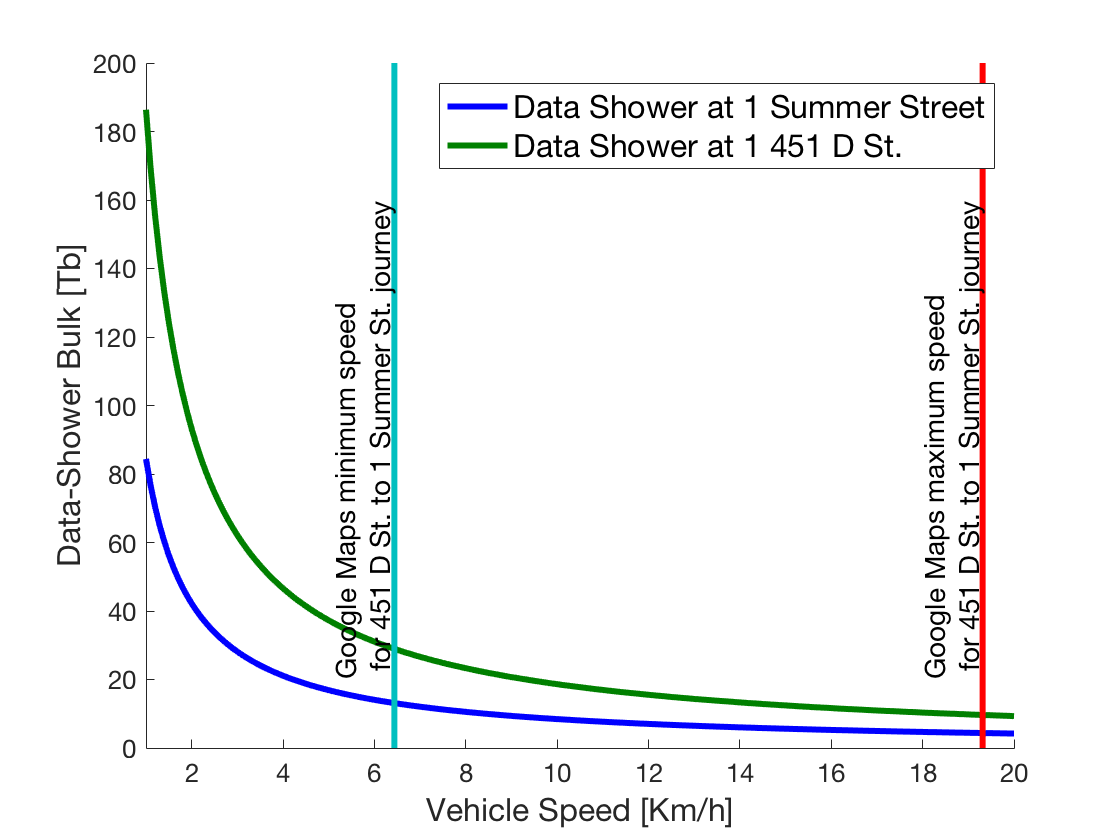}
\caption{Data Shower Bulk as a function of the average mule velocity. Single-way journey between two towers located at \textit{451 D St.} and \textit{1 Summer Street} and owned by \textit{Markley Group LLC} and \textit{XO Communications}, respectively, through the route suggested by Google Maps.}
\label{fig:m}
\end{minipage}
\end{figure*}

\subsection{Channel Modeling}
\label{sec:5.2}

The values for all the relevant parameters, used in this section, are summarized in Table~\ref{tab:1}. Their values are set according to previous works \cite{RappCh14,RappTWC15} and \cite{JorAky11,Han15}, as detailed below.

Regarding the mmWave communications, to provide a realistic estimation of the channel capacity, we use the experimental values of the mmWave channel parameters measured in \cite{RappCh14,RappTWC15} for both the LoS and NLoS propagation conditions, when the carrier frequency is $73$ GHz and the bandwidth is $1$ GHz.
As expected, the path loss for NLoS propagation conditions is significantly higher than the one in LOS propagation conditions. For the sake of clarity, in Fig.~\ref{fig:f} we report the experimental values of the LoS, NLoS and outage probabilities given in equations \eqref{eq:p_los_mm}, \eqref{eq:p_nlos_mm} and \eqref{eq:p_out_mm}, respectively~\cite{RappCh14,RappTWC15}. According to the experiments, the probability of having an outage event is null for distances smaller than $150$m, but it increases up to 0.7 for distances around $200$m.

Regarding the THz communication, we adopt an accurate channel modeling by accounting for the  molecular absorption characterizing USA high latitude locations at sea level during summer available in the Hitran Database \cite{hitran} as done in the seminal work in \cite{JorAky11}.
Accordingly, the total path-loss $|H_{\text{THz}}(f,d)|^2$, given in \eqref{eq:THzchannel}, is a function of both the distance and the frequency. We account for some unique findings in relation the THz bands from the previous works \cite{JorAky11,Han15}, i.e., the path loss in the THz band not only depends on the transmission distance and the system frequency, but also on the composition of the transmission medium at a molecular level through $k(f)$. Specifically, we observe that: $i)$ the path loss increases with both the distance and the frequency; $ii)$ several peaks of attenuation can be observed due to the molecular absorption loss controlled by $k(f)$; $iii)$ the molecular absorption defines several transmission windows along the frequency scale with varying widths that are defined by the molecular composition of the medium. 
%
It is clear that the spectrum range [0.8-0.9] THz represents a suitable band for THz communications up to $10$ meters, and the results derived in the following assume the use of such a band. Clearly, larger bands can be exploited by adopting distance-based modulation techniques and the results derived in the following continue to hold.

In Fig.~\ref{fig:g}, we report the values of the LoS and outage probabilities for the THz communications obtained according to the model \eqref{eq:p_o_thz}. The simulation setting is as follows: the transmitted power is $0$dBm and the minimum SNR $\gamma_{\text{th-THz}}$ required for establishing the THz link is given by:
\begin{equation}
\gamma_{\text{th-THz}} = k \gamma_{\text{THz}}(10)
\end{equation}
with $k$ ranging from $0.1$ to $1$, i.e., with $\gamma_{\text{th-THz}}$ being a fraction of the average SNR measured at a distance equal to $10$m.

The rationale for this model is twofold: $i)$ it allows us to abstract from the particulars of the THz transceiver, such as its sensitivity or noise figure; ii) it sounds reasonable to assume that the minimum SNR required for establishing a THz link is related to the SNR measured at the maximum distance at which the THz link could be established. We observe that for a transmitted power of $0$dBm, at $10$m we measure an outage probability of roughly 0.6 for $k=1$.

\begin{figure*}[t]
\centering
\begin{minipage}[c]{0.48\textwidth}
\includegraphics[width=1\columnwidth]{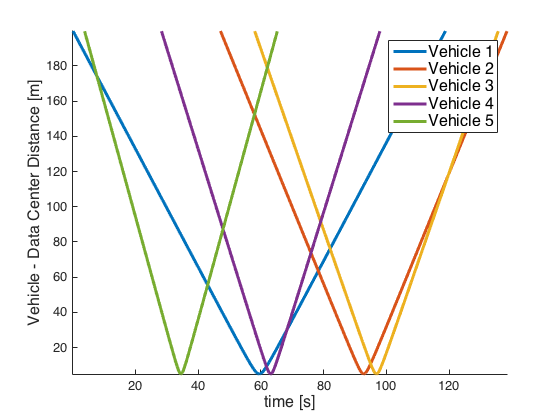}
\caption{Vehicle to data center distance as function of time for a single Monte Carlo realization. Minimum separation distance during closest approach is roughly 5 m. Used as input to generate Figure~\ref{fig:02}.}
\label{fig:01}
\end{minipage}
\hspace{0.02\textwidth}
\begin{minipage}[c]{0.48\textwidth}
\includegraphics[width=1\columnwidth]{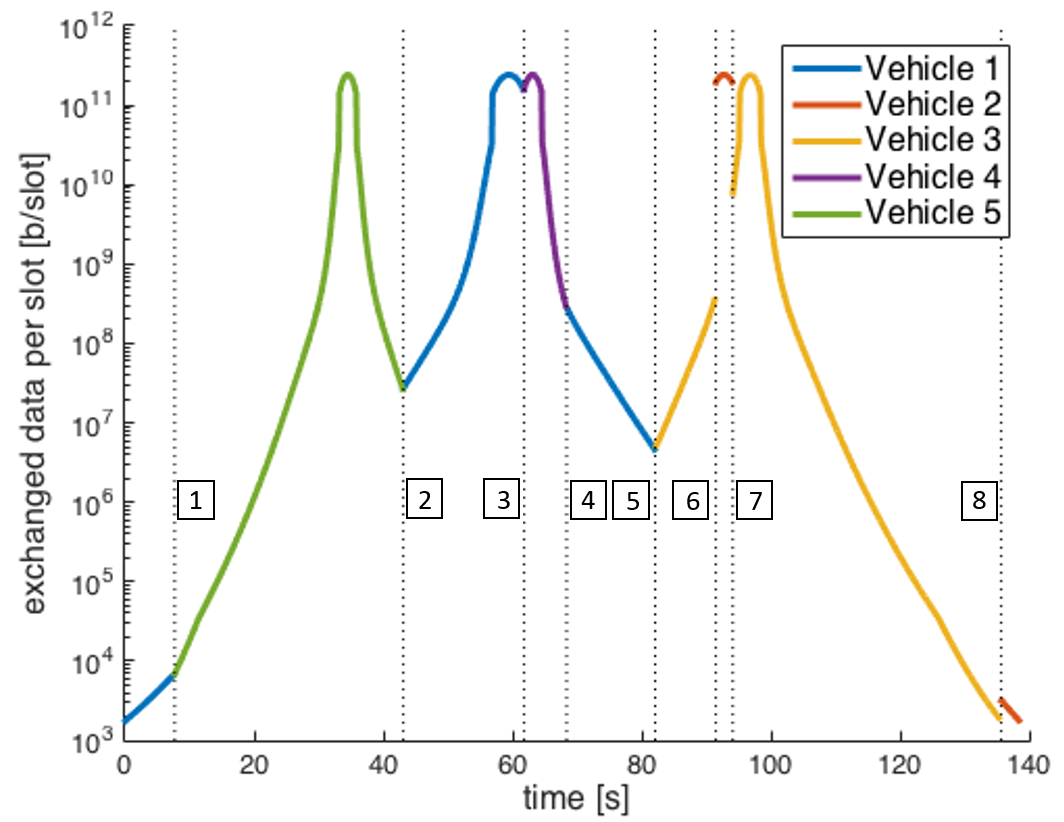}
\caption{Amount of exchanged data in every time slot by adopting the proposed greedy scheduling algorithm (Algorithm~\ref{alg:x.1}). Each switch is identified by the dotted vertical line with an associated index.}
\label{fig:02}
\end{minipage}
\end{figure*}
\begin{figure}[t]
\includegraphics[width=1\columnwidth]{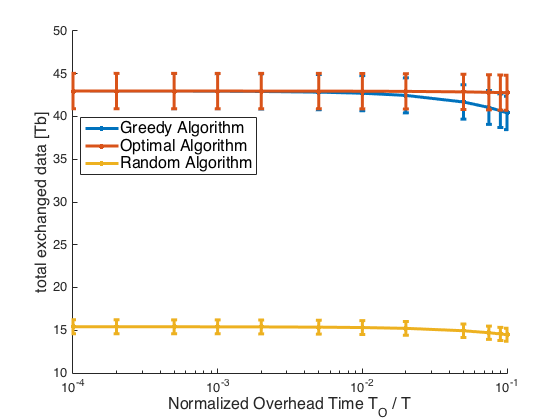}
\caption{Comparing the total exchanged data transferred with the three scheduling approaches.}
\label{fig:03}
\end{figure}
\subsection{Data Shower Performance Analysis}
\label{sec:5.3}

In Fig.~\ref{fig:h}, we report the THz capacity as function of the transmitted power and the separation distance between the transmitter and the receiver, by accounting for the outage loss as in \eqref{eq:p_o_thz}. Although we under-estimate the achievable THz capacity given in \eqref{eq:cap_thz} by limiting our attention to a single spectral window, we note that the achievable capacity is greater than $1$ Tbps for every values of the considered transmitted power at the maximum distance of $10$m. Furthermore, in presence of a LoS connection the achievable capacity at $10$m roughly increases of $1.5$ times.

In Fig.~\ref{fig:i}, we report the distance-dependent capacity available by adopting the proposed protocol, derived in \eqref{eq:cap_dist}, as a function of the distance and the THz transmitted power. Within the considered distance range $[1,200]$m, the achievable capacity varies of several orders of magnitude, ranging from Tbps to Mbps for distances around $200$m. This result is reasonable, since: i) in urban scenarios, the probability of a mmWave LoS connection decreases significantly as the distance increases, due to the building outage effects, and ii) mmWave NLoS path loss is particularly severe, with values exceeding $200$dB for distances greater than $100$m. Nevertheless, we note that the available capacity exceeds the Gbps and the Tbps for distances in the order of $10$ meters or less, respectively, for every value of the transmitted power.

In Figure~\ref{fig:l}, we show the data capacity derived in \eqref{eq:bit_tot} as a function of both the minimum distance $d_{\text{min}}$ between the transmitter and the receiver, and the average mule velocity. Specifically, we show that the amount of bits that can be transferred in a single-way journey between the vehicle, moving with constant-speed along a straight-trajectory, and the data center by adopting the proposed THz/mmWave mode selection. For a fair comparison, we assume that the transmitted powers of the mmWave and the THz links differs by at least 10dBm, i.e., we assume a Tx power of $30$dBm and $20$dBm for the mmWave and THz links, respectively. We adopt the same transmitter power value for mmWave communications used in the real-world experiments described in \cite{RappCh14}. For THz communications, we consider levels up to $20$dBm to account for the latest results experimentally achieved in submillimeter literature~\cite{SilesSubmmW, Leong085THz}. We note that the data rate increases as the average velocity decreases, having the mule spending more time in the range in which a mmWave/THz communication is possible. Hence, by controlling the velocity of the mule, an impressive transfer of information can be easily achieved. In particular, we observe that at the reasonable minimum distance of $4$m, we are able to transfer an amount of information exceeding one Terabit with a single journey in the worst case, i.e., when the average mule velocity is $10$Km/h. Even more impressive, when the average velocity is roughly $2$ Km/h, the amount of information exceeds 100 Terabit with a single journey for every considered minimum distance. These results suggest that by using the proposed mmWave/THz switching protocol, we can exchange much higher amount of data compared to what can be achieved with classical wired or wireless technologies.

In Fig.~\ref{fig:m}, we quantify the data transfer volume as function of the mule velocity for the real journey traced in Fig.~\ref{fig:b}, with the  mule reaching the (existing) tower located \textit{451 D St.} starting from the (existing) tower located at \textit{1 Summer Street}. The distance between the mule and data center as a function of time has been obtained from the journey route suggested by Google Maps. Specifically, the minimum distances between the transmitter and the receiver are $5.02$ and $5.03$m, respectively. The minimum and the maximum average speed, obtained through the Google Map estimation of the journey time are reported within the figure. We observe that the results shown in Fig.~\ref{fig:m} confirms that the data exchange of around 100 Terabit is possible with a single journey for each data center.  
\subsection{Scheduling Performance Analysis}
\label{sec:5.4}
To assess the performance of the proposed scheduling procedure, we consider multiple vehicles traveling along a two-lane road with constant velocity by first approaching and subsequently moving away from the Infrastructure.

The closest distance of approach to the data center is around 5 m as in Figure~\ref{fig:m}, and the vehicles enter within the mmWave operational range at random times. The vehicle velocities are picked uniformly at random within the range [3-7]m/s, modeling so a typical urban scenario. Hence, the vehicles are characterized by different contact times.

In the first experiment, we consider 5 vehicles to be scheduled. Fig.~\ref{fig:01} shows the distances between the vehicles and the data center as a function of time for a single Monte Carlo run. The time horizon is roughly two minutes, corresponding to 1387 slots. Given the variable arrival and contact times, the naive approach of scheduling vehicles on first-come-first-serve basis (even with all the other settings held identical) will clearly be sub-optimal.

This is confirmed by Fig.~\ref{fig:02}, which shows the amount of data exchanged by each vehicle in each time slot by adopting the greedy scheduling algorithm given in Algorithm~\ref{alg:x.1}. Fig.~\ref{fig:02} is obtained by considering the same Monte Carlo realization depicted in Fig.~\ref{fig:01} and with $D_{v}$ uniformly distributed in $[5, 15]$ Tb. A switch between scheduled vehicles happens at the time instants depicted with the dotted vertical lines. A total of 8 switches is observed in the entire time horizon. The contiguous set of slots assigned to a vehicle can be identified by the color. We note that three vehicles, i.e., vehicle 1, 2 and 3, are served in non-contiguous time slots so that the overall throughput can be maximized. Clearly, as pointed out in Sec.~\ref{sec:4b}, the greedy algorithm does not assure always the optimal solution, since it does not account for the time overhead in vehicle scheduling. Finally, we note that the peaks in the figure are indicative of the time slots during which the vehicles are in the THz operational range.

To substantiate the performance of the proposed greedy algorithm, in Fig.~\ref{fig:03} we compare it with the random scheduling and the optimal scheduling designed in Section~\ref{sec:4b}. Specifically, we implement the optimal scheduling through the exhaustive search of the solution maximizing the total number $N$ of exchanged bits, among the admissible solutions satisfying constraints \eqref{eq:x.5} and \eqref{eq:x.6}. For a fair comparison, we implement the random scheduling by selecting uniformly at random one solution among the admissible ones. To assure practical time and memory complexity for the optimal and the random algorithms, we limit the number of vehicles to two.

Fig.~\ref{fig:03} shows the average total exchanged data as a function of the normalized overhead time, i.e., $T_O / T$, for $1000$ Monte Carlo runs. The vertical bars denote the $95\%$ confidence intervals. The relative performances of three algorithms in comparison is quite apparent. Importantly, the greedy algorithm exhibits excellent performance overall and an optimal performance for overhead time lower than $10^{-2} T$.

Intuitively, this can reasoned in the following way: Greedily assigning the set of vehicles that can complete its data transfer needs at the earliest, and subsequently removing those vehicles that have successfully completed the data exchange from any further assignment is a clever strategy for it rapidly makes progress in reducing both the overall backlogged data from all vehicles, and the overhead time that will be needed in the future within the time horizon. Since the greedy approach prioritizes the completion of data needs of each vehicle over the overhead cost, the performance deteriorates slightly from that of the exhaustive search for large overhead time relative to the slot time.

Not surprisingly, the random algorithm performs poorly as it disregards the variable amount of backlogged data on each vehicle. Moreover, random assignment on every slot implies a poor utilization of high bandwidth as it essentially keeps switching across vehicles and accumulates all the coordination overhead time. This impairment in performance becomes all the more apparent and severe when the normalized overhead time approaches unity -- the overhead time occupies the entire slot time thereby leaving no time for data exchange.

\section{Conclusion}
\label{sec:6}
We developed a handoff and medium access protocol that allows vehicles to dynamically switch between the mmWave and THz links for high bandwidth data transfer operations. We derived the capacity of the network that results from the protocol operation, and demonstrated how the switching action between these two access methods results in significant improvements over a single and constant choice. Furthermore, we proposed an optimal procedure at the SDN controller for scheduling multiple vehicles for accessing a given small cell tower. Since the search for the optimal scheduling is a NP-hard problem, we design a computational-feasible greedy scheduling algorithm, exhibiting a polynomial-time complexity and excellent performance with respect to the optimal scheduling algorithm. Finally, we quantified the actual end to end data transfer rates possible for two tower locations within the Boston area. The analysis showed that a transfer of around $100$ Terabit is possible with a single journey, by controlling the velocity of the mule.

\appendices
\section{Proof of Proposition~\ref{prop_1} }
\label{app_1}
To prove the proposition, first, we note that, due to the relative movement, the distance between the mule and the SD-BS is a function $f(\cdot)$ of the time, whose expression depends on the mule mobility patterns, i.e., $d(t)=f(t)$. 
Since the closed-form expression for the channel capacity derived in \eqref{eq:cap_dist} is a function of the distance, for each time $t \in [t_{\text{in}}, t_{\text{out}}]$, the relative distance $d(t)$ has to be computed in order to evaluate the corresponding capacity. 
By accounting for this, the proof easily follows by observing that, 
according to the proposed protocol, $i)$ the transmitter and the receiver can exchange data only if their relative distance $d(t)$ at a certain time $t \in [t_{\text{in}}, t_{\text{out}}]$ is smaller than $d_{\text{th}}^{\text{mm}}$; $ii)$ when at a certain time $t \in [t_{\text{in}}, t_{\text{out}}]$ the distance $d(t)$ is $d_{\text{th}}^{\text{THz}} < d(t) \leq d_{\text{th}}^{\text{mm}}$, the communication in mmWave band is not affected by a reduction of the available capacity for transmitting control packets, since the LTE interface is used for this purpose (i.e., to return ACKs); $iii)$ when at a certain time $t \in [t_{\text{in}}, t_{\text{out}}]$ the distance $d(t)$ is $0 < d(t) \leq d_{\text{th}}^{\text{THz}}$, the communication in THz band does not suffer a reduction of the available capacity for transmitting control packets, since the mmWave interface is used for ACKs. Hence, all the times $t \in [t_{\text{in}}, t_{\text{out}}]$ such that the corresponding distances $\{d(t)\}$ are smaller than $d_{\text{th}}^{\text{mm}}$, i.e., for which the transmitter and the receiver are in contact, may be ideally dedicated for data transfer.

\section{Proof of Corollary~\ref{cor_1}}
\label{app_2}
The proof easily follows by accounting for the result in Proposition~\ref{prop_1} as well as the hypothesis of uniform strict movement. Specifically, we first observe that if the mule is traveling along the path between the point A and E, as depicted in Fig.~\ref{fig:updl}, according to a uniform strict movement, at each time $t$ belonging to $[t_{\text{in}}, t_0]$ the corresponding distance $d(t)$ belong to $\mathcal{R}_{mm} \bigcup \mathcal{R}_{THz}$. In addition, since the velocity is uniform, all the distances covered during such a movement contribute equally to the computation of the overall capacity $\mathcal{C}^{\mathcal{O}}$ that, as a consequence, can be computed as:
\begin{align}
\label{eq:cap_tot}
\mathcal{C}^{\mathcal{O}}=\frac{1}{{d_{\text{th}}^{\text{mm}}-d_{\text{min}}}}\displaystyle\int_{d_{\text{min}}}^{d_{\text{th}}^{\text{mm}}}\displaystyle\mathcal{C}(\eta) d\eta
\end{align}
Then we observe that, to compute the net transferred bits, it is sufficient to multiply such an overall capacity with the average time $T_c$ in which the transmitter and the receiver are in contact, i.e., the time spent to travel the path from A to E. In fact, as observed in the proof of Proposition~\ref{prop_1}, since the ACKs are sent through the second-best available option, there is no reduction in the available capacity for transmitting control packets. Hence, the entire $T_c$ may be ideally dedicated to data transfer.\\
However, such a contact time has to be reduced by $\epsilon_{s}^{\text{mm}}$ to account for the time spent at the start of the mmWave communication to synchronize the transmitter and the receiver at a finer granular level. In fact, as detailed in Section~\ref{sec:2}, although the transmitter and the receiver know the positions of each others, around point A in Fig.~\ref{fig:updl}, a granular synchronization is needed. Similarly, around point B in Fig.~\ref{fig:updl}, an additional time $\epsilon_{s}^{\text{THz}}$ is devoted for synchronizing the transmitter and the receiver at a finer granular level to start the THz communications. 
As detailed in Section~\ref{sec:3}, the interaction time of the vehicle with the data center is divided into distinct uplink (UL) followed by downlink (DL) phases. This implies that the contact time has to be reduced by an additional quantity $\epsilon_{tr}$ to account for the switching time transceivers spend to change their operational mode. Hence, by accounting for the above analysis we get that the time available for data transmission is given by:   
\begin{align}
\label{eq:app_1_1}
T= T_c- \epsilon_{s}^{\text{mm}}-\epsilon_{s}^{\text{THz}}-\epsilon_{tr}
\end{align}
By multiplying \eqref{eq:app_1_1} with \eqref{eq:cap_tot}, the proof easily follows by further observing that, under the hypothesis of uniform strict movement, $T_c$ is given by $\frac{2 d_{\text{th}}^{\text{mm}} \cos \alpha}{v}$, where $v$ is the uniform average velocity and $\alpha$ is the angle formed by: (i) the distance $d_{\text{th}}^{\text{mm}}$ between the mule and the tower at time $t_{\text{in}}$, and (ii) the direction of the movement. 

\section{Proof of Proposition~\ref{prop:x.1}}
\label{app_prop_scheduling}

By accounting for constraint \eqref{eq:x.6}, at most one vehicle can be scheduled by the data center in each time slot. Hence, the average number of bits $N_v$ exchanged with vehicle $v$ during the entire time horizon $\mathcal{K}=\{k_i,\ldots,k_e\}$ is obtained as sum of bits $n^k_v$ exchangeable in time slot $k$, for each time slot $k$ assigned to vehicle $v$, i.e., $N_v = \displaystyle \sum_{k \in \mathcal{K}} \phi^k_v n_v^k$.

When two consecutive time slots are assigned to different vehicles, a scheduling overhead cost has to be paid. To account for such an event, we define the indicator function $\chi^k_v$ as in \eqref{eq:x.10}, and by exploiting the equation~\eqref{eq:cap_dist}, it results:
\begin{equation}
\label{app_x.8}
n_v^k = \int_{(k-1) T + \chi^k_v T^O_v}^{kT} C_v(d_v(t)) dt.
\end{equation}
The proof easily follows by noting that the total number $N$ of exchanged bits in the considered time horizon $\mathcal{K}$ is given by the sum of the average numbers of bits $\{N_v\}$ exchanged with the vehicles in $\cup \mathcal{V}_k$

\bibliographystyle{IEEEtran}    
\bibliography{mybibfile,ThzmmWbio,THz}

\end{document}